\documentclass[pra,twocolumn,showpacs,preprintnumbers,float,amsmath,amssymb]{revtex4-1}
\usepackage{graphicx}% Include figure files
\usepackage{dcolumn}% Align table columns on decimal point
\usepackage{bm}% bold math
\def \ed {\end{document}}
\def\Fbox#1{\vskip1ex\hbox to 8.5cm{\hfil\fboxsep0.3cm\fbox{%
\parbox{8.0cm}{#1}}\hfil}\vskip1ex\noindent}  %%  {TEXT} in BOX
\newcommand{\eq}[1]{(\ref{#1})}%%  requires \eq{label}
\newcommand{\Eq}[1]{Eq.~(\ref{#1})}%%  requires \eq{label}
\newcommand{\Eqs}[1]{Eqs.~(\ref{#1})}%%  requires \eq{label}
\newcommand{\Fig}[1]{Fig.~\ref{#1}}%%  requires \Fef{label}
\newcommand{\Figs}[1]{Figs.~\ref{#1}}%%  requires \Fef{label}
\newcommand{\Sec}[1]{Sec.~\ref{#1}}%%  requires \Fef{label}
\newcommand{\Ref}[1]{Ref.~\cite{#1}}%%  requires \Fef{label}
\def\be{\begin{equation}}\def\ee{\end{equation}}
\def\bea{\begin{eqnarray}}\def\eea{\end{eqnarray}}
\def\bse{\begin{subequations}}\def\ese{\end{subequations}}
\newcommand{\BE}[1]{\begin{equation}\label{#1}}
\newcommand{\BEA}[1]{\begin{eqnarray}\label{#1}}
\newcommand{\BSE}[1]{\begin{subequations}\label{#1}}

\let \= \equiv \let\*\cdot \let\~\widetilde \let\^\widehat \let\-\overline
\let\p\partial
\def\<{\left\langle}    \def\>{\right\rangle}
\def\({\left(}          \def\){\right)}
\def \[ {\left [} \def \] {\right ]}

\newcommand{\B}[1]{{\bm{#1}}}%% Bold Roman & Greek Lower & Upper Case
\newcommand{\C}[1]{{\mathcal{#1}}}    %%   Calligrapfic Upper case
\renewcommand{\sb}[1]{_{\text {#1}}}  %% sub-   for lower case
\renewcommand{\sp}[1]{^{\text {#1}}}  %% super- for lower case
\begin{document}

\title{%{\rm \emph{DRAFT, not for distribution \hfill Version of \today, at 8:00 PM  }}\\
Direct Energy Cascade in Two-Dimensional Compressible Quantum Turbulence}

\author{Ryu Numasato and Makoto Tsubota}
\affiliation{Department of Physics, Osaka City University, Sumiyoshi-ku, Osaka 558-8585, Japan}%

%\author{Makoto Tsubota}
%\affiliation{Department of Physics, Osaka City University, Sumiyoshi-ku, Osaka 558-8585, Japan}%

\author{Victor S. L'vov}
\affiliation{Department of Chemical Physics, The Weizmann Institute of Science, Rehovot 76100, Israel}

\date{\today}% It is always \today, today,

\begin{abstract}
We numerically study two-dimensional quantum turbulence with a Gross--Pitaevskii model. With the energy initially accumulated at large scale, quantum turbulence with many quantized vortex points is generated. Due to the lack of enstrophy conservation in this model, direct energy cascade with a Kolmogorov--Obukhov energy spectrum $E(k) \propto k^{-5/3}$ is observed, which is quite different from two-dimensional incompressible classical turbulence in the decaying case. A positive value for the energy flux guarantees a \emph{direct}   energy cascade   in the inertial range (from large  to small scales). After almost all the energy at the large scale cascades to the small scale, the compressible kinetic energy realizes the thermodynamic equilibrium state without quantized vortices.
\end{abstract}


\maketitle

\section{\label{s:intro} Introduction}
The experimental discovery of Bose--Einstein condensates (BEC) 15 years ago~\cite{And-95,Dav-95,Bra-95}, long after their theoretical prediction in 1924--25 \cite{Bose-24,Ein-25}, has renewed interest in this field. BEC systems are of great interest as they
promise the opportunity to study new nonlinear dynamical systems built
with a high degree of control and flexibility. Also, theoretical and numerical studies of BEC
systems are  of  general importance for nonlinear physics, as BEC systems are described by one of the most important and universal partial differential equations, the nonlinear Schr\"odinger equation, called in this field the Gross--Pitaevskii (GP) equation (GPE)~\cite{Pit-03,Peth-02}:
\begin{equation}\label{GPE}
i \frac{\partial \Psi}{\partial t}+\frac 12 \B\nabla^2 \Psi - \frac 12 g |\Psi|^2 \Psi = 0\ .
\end{equation}
Here  the condensate wave function $\Psi(\bm r, t)$  plays the role of  the complex order parameter.

The GPE~(\ref{GPE}) describes a Bose gas at low temperature, which  may behave similarly to a superfluid, and
can describe its random (turbulent) motion, i.e.\ quantum turbulence (QT).
QT physics, comprising
tangled quantized vortices, is an important
research topic in low-temperature physics~\cite{HT-2008,Tsu-08}. Stimulated
by recent experiments on both superfluid $^3$He and superfluid
$^4$He, where a few similarities have been observed between
quantum and classical turbulence~\cite{Nor-97,Abi-03,Par-05,Tsu-05,Tsu-06,Hel-front,Mon-cube,Our-rev}, studies on QT have entered a new stage where one of the main motivations
is to investigate the relationship between quantum and classical turbulence.
In particular, the Kolmogorov--Obukhov turbulent kinetic energy spectrum $E(k)\propto k^{-5/3}$ has been observed in laboratory experiments on superfluid $^4$He similar to that in normal fluids~\cite{MT-98}.  The physical explanation for this is very simple: for superfluid motion at scales $\C L$ essentially exceeding the mean intervortex distance $\ell$, the quantization of the vortex lines can be neglected. This range of scales corresponds to the
quasi-classical limit and  this type of turbulence can be called \emph{quasi-classical turbulence} (QCT). QCT in the three-dimensional (3D) case has subsequently been found in numerical simulations with the GPE  as well as in the vortex filament model~\cite{ATN-02} that can describe  motions  of quantized vortex lines with a prescribed   core structure.

It is well known that Euler equations, that describe motion of ideal (inviscid) fluids, in the two-dimensional (2D) case exhibit an additional second quadratic motion invariant, enstrophy $\Omega$, as well as kinetic energy $E$:

\BE{enst}
E\= \frac 12 \int  |\B v| ^2 \, d \B r\,,\quad
\Omega \= \int  \omega ^2 \, d \B r\,,  \quad
\omega\= [ \B \nabla \times \B v ]_z\ .
\ee
Kraichnan recognized~\cite{Kra-67} that the appearance of $\Omega$ drastically modifies the physics of 2D turbulence. It changes the direction of the energy flux in the classical Richardson--Kolmogorov cascade. Instead of a \emph{direct} energy cascade (from large to small scales), like in 3D turbulence, in the 2D case there is an \emph{inverse} energy cascade, from small to large scales. In both the 3D- and 2D-cases the energy spectrum is the same,
$E(k)\propto k^{-5/3}$. In 2D turbulence the  direct  \emph{energy} cascade is replaced by the direct enstrophy cascade with the energy spectrum $E(k)\propto k^{-3}$ (subject to certain logarithmic corrections, unimportant in the current discussion).  These predictions have been confirmed in laboratory experiments, see e.g. Ref.~\cite{Tab-2D-review},  and large scale direct numerical simulations (DNS) of the Navier--Stokes equations, see e.g. Refs~\cite{2D-NSE-DNS-1,2D-NSE-DNS-2,2D-NSE-DNS-3}.

An important issue is whether the inverse energy cascade is a feature of 2D QT. The qualitative answer given in our paper is ``not necessarily". The physical reason for this is quite simple and general.

In superfluids, the enstrophy $\Omega$ coincides (up to a prefactor) with the total number of quantized vortex points \cite{Ryu-10}\label{Ryu}. In GP dynamics the total number of vortices is not conserved. They can appear in pair-creation or disappear in pair-annihilation. Therefore, Kraichnan's arguments~\cite{Kra-67} become, generally speaking irrelevant for 2D QT. Thus, for a certain range of parameters, where the creation/annihilation of vortex pairs become dynamically important, we can expect a \emph{direct} energy cascade in 2D QT, exactly as in 3D classical turbulence. This is an important  conclusion of our paper.   Simulating the free evolution of 2D GPE (from an initial condition with energy located at large scales) we observed a \emph{direct} energy cascade with the Kolmogorov--Obukhov energy spectrum $E(k)\propto k^{-5/3}$.

The same argument can be made based on Euler equations. Indeed, the superfluid density in the vicinity of a vortex core is different from that in a vortex-free superfluid. Therefore the creation/annihilation of vortex pairs lead to variations of the superfluid density. In other words,  the GPE is reduced to a compressible Euler equation. However, a compressible Euler equation does not preserve enstrophy. Therefore at some level of compressibility (characterized by the Mach number, the ratio of the turbulent velocity fluctuations to the sound velocity) the direction of the energy flux can change sign and instead of an inverse energy cascade, we observe a direct cascade, typical for 3D turbulence.

In this study, we  observe  the  direct energy cascade in numerical experiments with 2D GPE.  Starting from initial conditions with energy localized at  large scales we observed intermediate asymptotic Richardson cascade transporting energy of the system toward small scales. The energy spectrum  at this stage is close to the Kolmogorov--Obukhov law $E(k)\propto k^{-5/3}$. At later stages, when an essential part of the  energy reaches the smallest scales available in our simulations (mean intervortex spacing, which is a few times larger than the core diameter), the system quickly evolves  toward thermodynamic equilibrium. As an independent test of the direction of the energy flux we analyzed the energy balance equation in $k$-space and found that the sign of the energy flux indeed corresponds to the direct cascade.

In addition, we studied the properties of the finite steady-state of the system by a different means, e.g. by analyzing the power spectra in the frequency domain for motion with different wave vectors. We show that the position of the maximum of the power spectra $\omega \sb {max}(k)$ depends on the dispersion relation derived from Bogoliubov's microscopic theory\cite{Pit-03, Peth-02}.

In addition, we investigate how the time evolution of the
system depends on the level of nonlinearity by qualitatively comparing
its behavior with the same initial conditions but
 different coupling constants $g$.

The paper is organized as follows. After this introduction,
in Sec.\ref{s:model} 
we explain the analytical and numerical formulations of the GP model of compressible quantum turbulence. 
Next, in Sec.~\ref{s:GPE}
%\fbox{pls check this \#} 
we explain the turbulent state, i.e. the
time development of the energy, kinetic energy spectra, and
incompressible kinetic energy flux. We also show how the
direct energy cascade with the Kolmogorov--Obukhov law is
formed and prove the direct energy cascade with the
use of the approximated incompressible kinetic energy flux.
Then, through numerical analysis of the power spectrum
of the compressible effective velocity, we find that Bogoliubov's
microscopic theory holds in the thermodynamic
equilibrium state full of compressible kinetic energy.
Finally, in Sec \ref{s:concl} we state the conclusions of the paper.

\section{\label{s:model} GP model of compressible turbulence}

\subsection{\label{ss:GPvsEuler}GP and ``quantum" Euler equations}
It is well known~\cite{Pit-03,Peth-02} that GPE conserves the total energy (Hamiltonian) $H$ and ``particle number" $N$:
\BSE{conGPE}
\begin{eqnarray}\label{H}
H &=&  \int \Big [\frac 12 |\B\nabla \Psi|^2   + \frac 14 g |\Psi|^4\Big ] \, d \B r\,,  \\
N &=& \int  | \Psi|^2\, d \B r\ . \label{N}
\end{eqnarray}
\ese
By analogy with the Schr\"odinger equation we can consider the probability $|\Psi(\B r,t)|^2$ as a particle density:
\BSE{HD}%%
\BE{rho}%%
\rho(\B r,t)\=|\Psi(\B r,t)|^2\ .
\ee%%
Then the conservation law of $N$ can be considered as the conservation of the total mass of the system   $M\=\int \rho \, d \B r=   N$ (assuming that the ``particle mass" is unity), which can be written as  a continuity equation:
\BE{cont}
\frac{\p \rho}{\p t}  +  \B \nabla \cdot \B j=0\ .%%
\ee%%
Here  $\B j$ is the particle flux\:
\BE{flux}%%
\B j \=  \frac i2    \big ( \Psi  \B \nabla \Psi^*  -  \Psi^* \B \nabla \Psi \big ) \,,%%
\ee%%
which can be presented in the familiar form $\B j= \rho\,  \B v$  with the ``fluid" velocity:
\BE{vel}
\B v = \B \nabla \theta\ .
\ee
Here the phase $\theta$ is defined via    the Madelung  transformation:
\BE{MT}
\Psi= \sqrt {\rho}\exp \, (i\theta)\,,
\ee\ese%%
which maps the GPE~(\ref{GPE}) to the Euler equation for an ideal compressible fluid
with an extra quantum pressure  term.

\subsection{\label{ss:E-dec}Decomposition of the system energy}
Following~\Ref{Nor-97}   it is instructive   to decompose the total energy of the system $H$, \Eq{H},  conserved by GPE~\eq{GPE}, into four parts. The first part originates from the interaction energy in the Hamiltonian~\eq{H} of the GPE~\eq{GPE} with the density $\displaystyle g|\Psi|^4/4= g \rho^2/4 $. Bearing in mind that $\int \rho\,  d \B r=$const., we can introduce the density of the ``internal energy" as $ g\, \rho^2/4$ counted from $(g/2)\, (\rho - \frac 12)$:

\BE{E-int}
\C E\sb {int} (\B r)\= \frac  {g}4 (\rho-1)^2\ .
\ee

Next, we define the density of the fluid kinetic energy  as usual:
\BSE{E-dec}
\BE{E-kin} \C E\sb {kin}(\B r)\= \frac12 \rho |\B v|^2\= \frac 12 |\B w|^2\,, \quad \B w\= \sqrt \rho \, \B v\ .
\ee
This kinetic energy  can be divided into compressible and incompressible parts by decomposing the effective ``velocity" field $\B w$ into divergent free $\B w\sp i$ and potential $\B w\sp c$ parts:
\BEA{E-decA} \C E\sb {kin}(\B r)&=&\C E\sb {kin}\sp i(\B r)+\C E\sb {kin}\sp c(\B r)\,,\\
\C E\sb {kin}\sp i (\B r) &\=& \frac12  \,  |\B w\sp i|^2\,, \quad  \C E\sb {kin}\sp c (\B r) \= \frac12  \,  |\B w\sp c|^2\,,
\eea\ese
because $ \B w\sp i\cdot \B w\sp c\=0$.

Now, the first term in the Hamiltonian~\eq{H}  in the representation of~\eq{vel},~\eq{MT} can be presented as follows:
\bea \nonumber
\frac 12 |\B \nabla \Psi|^2&=& \frac 12 \Big [ \rho |\B v |^2  +  |\B \nabla \sqrt {\rho}|^2 \Big ]= \C E\sb {kin}(\B r)+ \C E \sb {qnt}(\B r) \\ \label{div}
\C E \sb {qnt}(\B r) &\=& \frac 12 \, |\B \nabla \sqrt {\rho}|^2\ .
\eea
The second term here, $\C E \sb {qnt}(\B r)$,  has been termed the \emph{quantum}  energy density~\cite{Abi-03}.
% (in physical space).
%%

%%

Finally,  the total (conserved) energy of the system can be decomposed into four parts:
\BE{en} H=\int  \Big[\C E \sb {int}(\B r)+ \C E\sb {kin}\sp i (\B r) + \C E\sb {kin}\sp c (\B r)+ \C E \sb {qnt}(\B r) \Big ]d \B r \ .
\ee

\subsection{\label{ss:num} Numerical procedure}
We solve  the 2D GPE~\eq{GPE}  by the pseudo-spectral method in the domain $256^2$ with  the fourth-order Runge--Kutta method for the time development. For details, see Ref.~\cite{Tsu-05}.

To generate large-scale turbulent flow, the initial condition of the wave function is set to the random phase state $\Psi(\bm r, 0)=\exp [{i\theta (\bm r,0)}]$ with
\begin{eqnarray}\label{theta}
\tilde{\theta }(\bm k,0)=
\left\{
 \begin{array}{ll}
    \theta  _0 \exp [i\alpha (\bm k)] & (\Delta k \le |\bm k| \le 3\Delta k)\,, \\
    0 & (\rm{otherwise}) \ .
  \end{array}
\right.
\end{eqnarray}
Here $\tilde{\theta }(\bm k,0)=\tilde{\theta }^{\ast}(-\bm k,0)$,
$\alpha(\bm k)$ is randomly taken for each $\B k$ in the range $(-\pi,\pi)$, and $\Delta k=2\pi/L$
is the wavenumber grid. The initial density is uniform,  $|\Psi(\bm r, 0)|^2=1$.
The phase distribution~(\ref{theta}) is energetically high and unstable so that many quantized vortex pairs are created.
With these initial conditions  we obtain, during time evolution, a 2D QT composed of a random configuration of quantized vortices.

\begin{figure*}[t]
\includegraphics[width= 0.99 \textwidth]{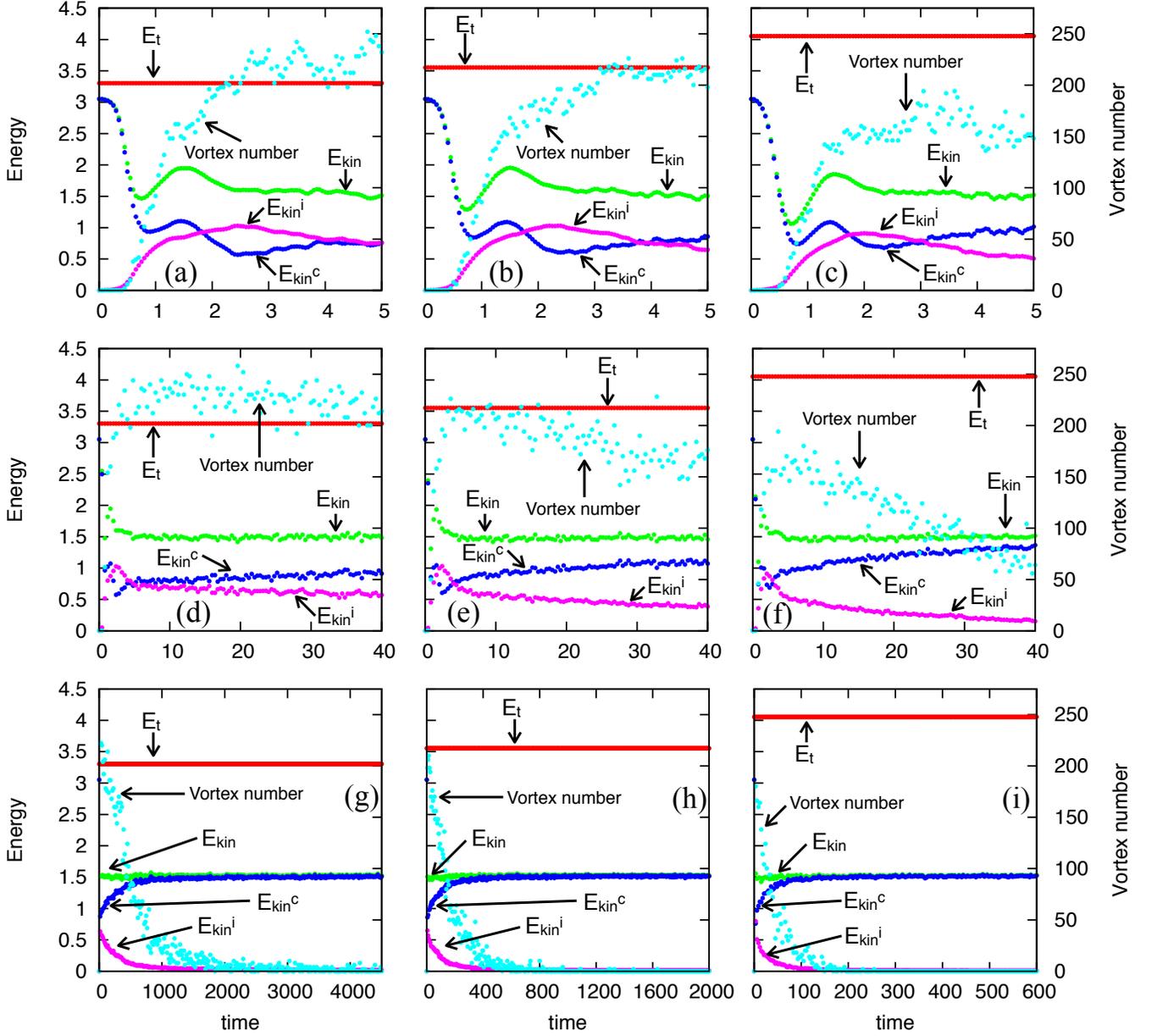}
\caption{\label{f:1} (Color online) Time evolution of the dimensionless total energy $E\sb t$, total kinetic energy $E\sb {kin}$ and its compressible $E\sb {kin}\sp c$ and  incompressible $E\sb {kin}\sp i$ parts. All energies are normalized with the particle number $N$, as shown in \Eqs{spec}. The time dependence of the total vortex  number (a measure of the enstrophy) is also shown. Left panels: $g=1$, Middle panels: $g=2$, Right panels: $g=4$. Upper panels: initial evolution, $t \le 5$, Middle panels: intermediate stage, $t \le 40$, Lower panels:  latest stage, $t \le 4500 $ for $g=1$,   $t \le 2000$ for $g=2$, and  $t \le 600$ for $g=4$.}
\end{figure*}

\begin{figure*}[t]
\includegraphics[width= 0.99 \textwidth]{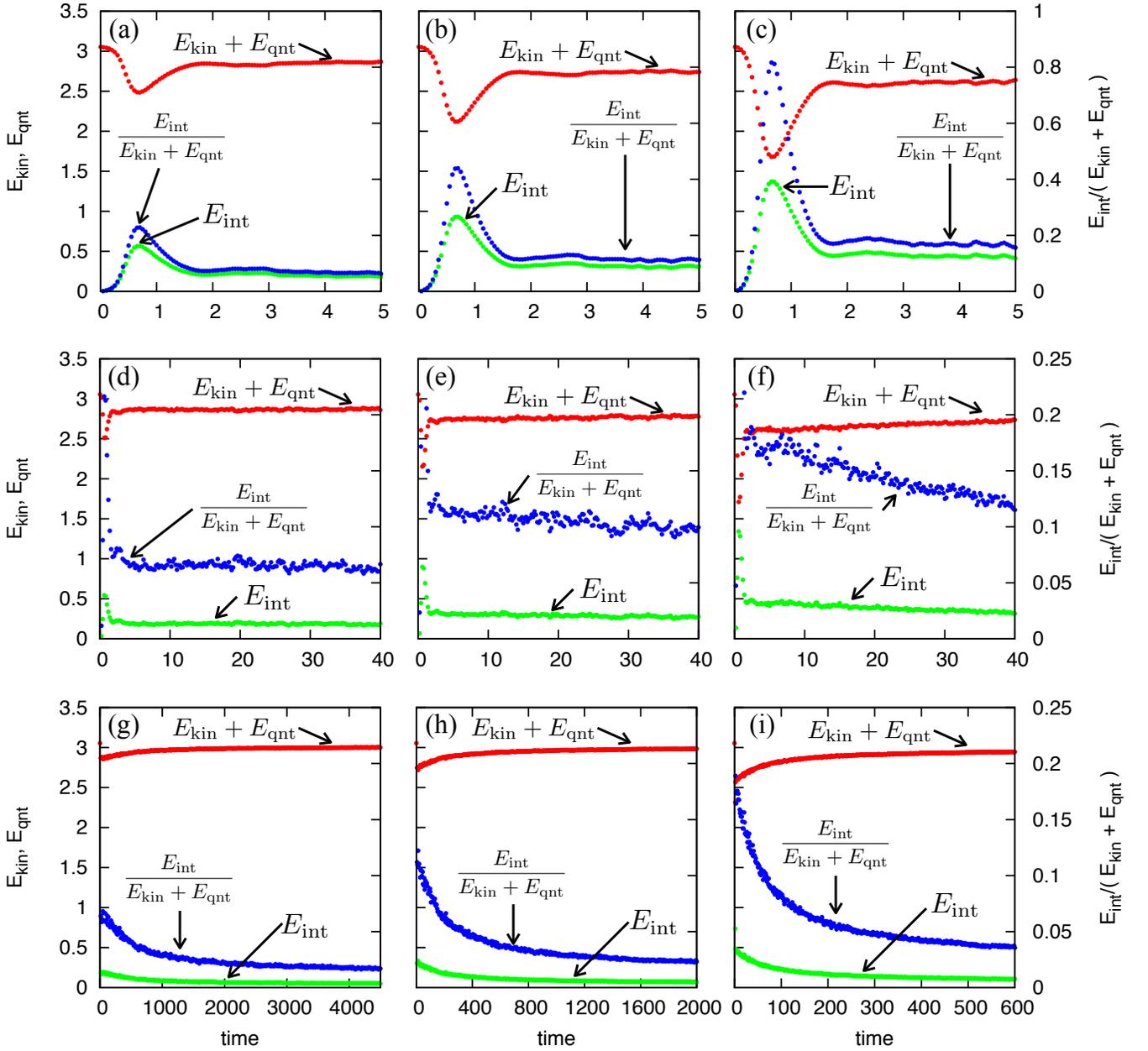}
\caption{\label{f:2} (Color online) Time evolution of the interaction energy $E\sb {int}$ (\Eq{E-int}), the sum of the total kinetic energy and quantum energies
(\Eqs{E-dec} and \eq{div}) and the dimensionless nonlinearity level, $R$ (\Eq{nl}).  All energies are normalized with the particle number $N$, as shown in \Eqs{spec}.  Left panels: $g=1$, Middle panels: $g=2$, Right panels: $g=4$. Upper panels: initial evolution, $t \le 5$, Middle panels: intermediate stage, $t \le 40$. Lower panels:  latest stage, $t \le 4500 $ for $g=1$, $t \le 2000$ for $g=2$, and $t \le 600$ for $g=4$.}
\end{figure*}

\begin{figure*}
\includegraphics[width= 0.99 \textwidth]{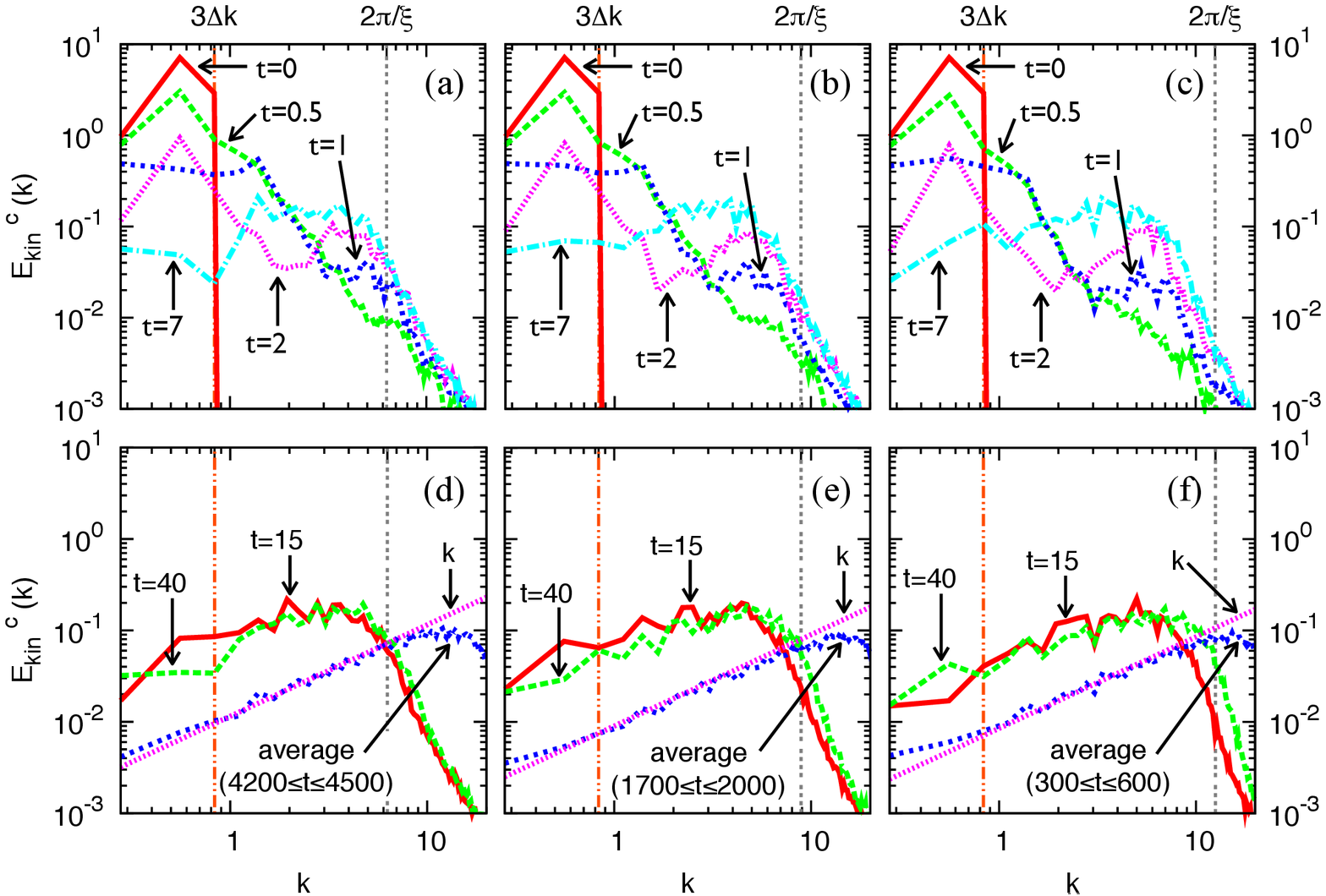}
\caption{\label{f:3} (Color online) Log--log plots
    of the spectra of compressible kinetic energy at earlier (upper panels) and later (lower panels) moments of time.  Left panels: $g=1$, Middle panels: $g=2$, Right panels: $g=4$. }
\end{figure*}

\begin{figure*}
\includegraphics[width= 0.99 \textwidth]{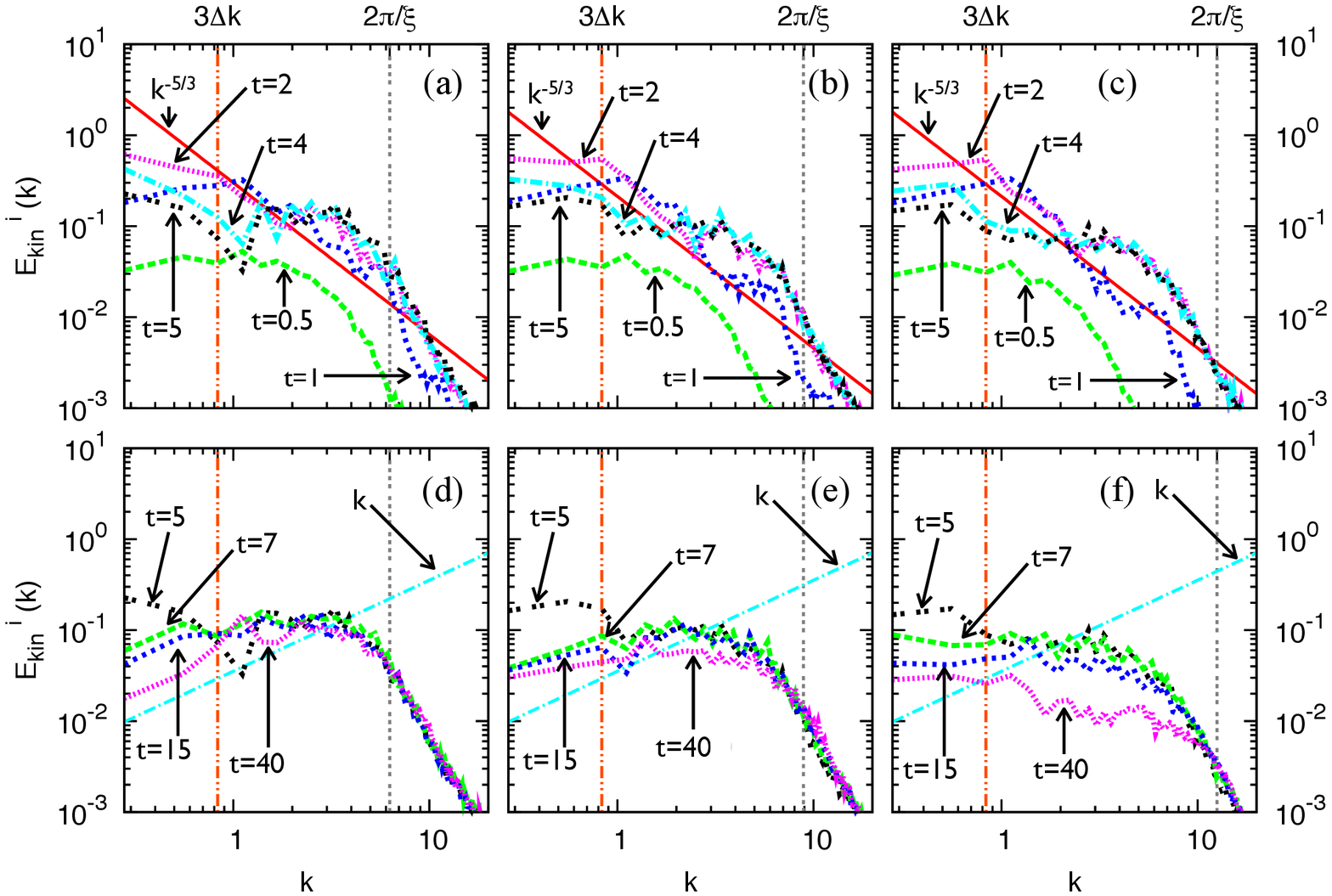}
\caption{\label{f:4} (Color online) Log--log plots
    of the spectra of incompressible kinetic energy at earlier (upper panels) and later (lower panels) moments of time.  Left panels: $g=1$, Middle panels: $g=2$, Right panels: $g=4$}
\end{figure*}

\begin{figure*}
\includegraphics[width= 0.99 \textwidth]{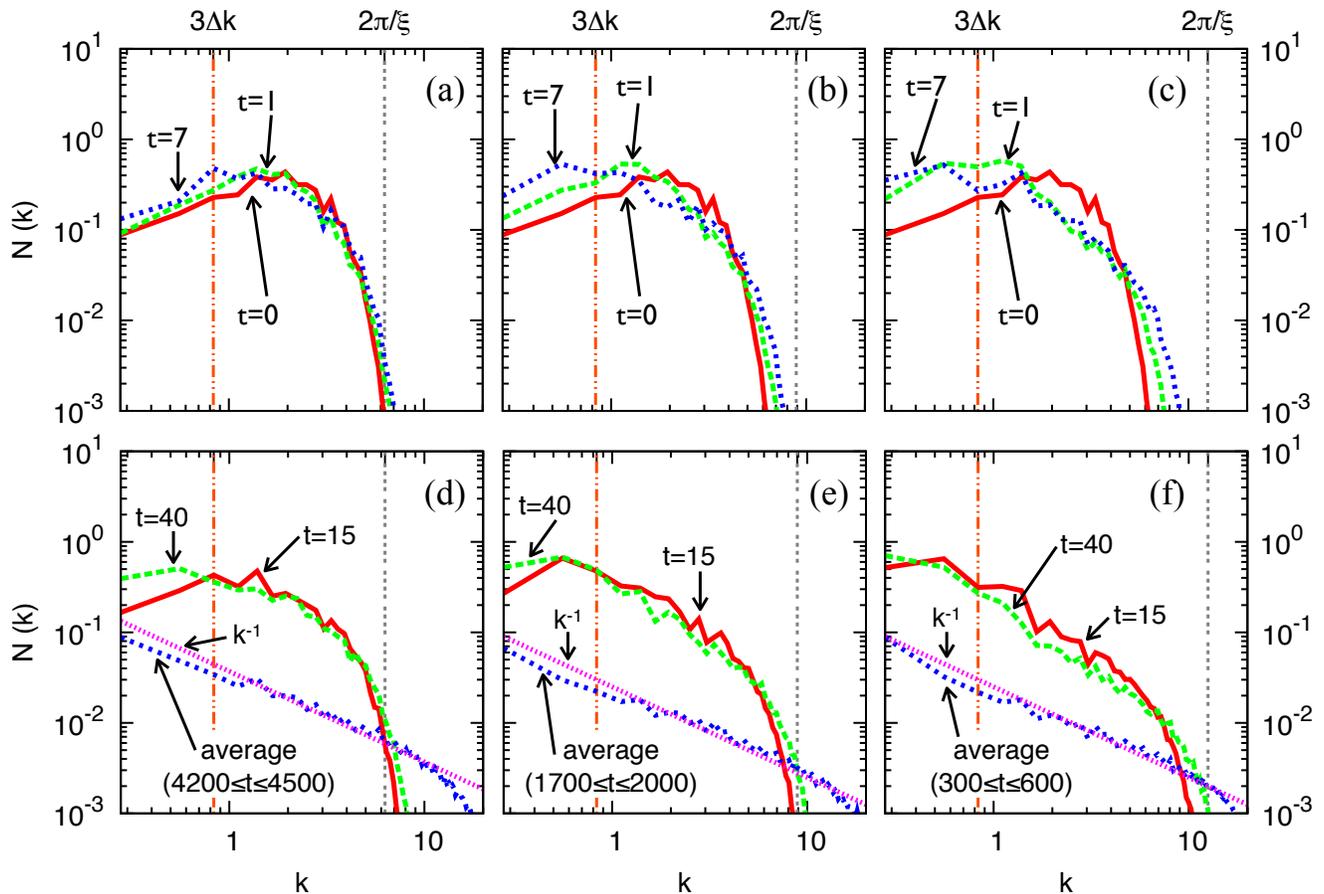}

\caption{\label{f:5} (Color online) Log--log plots
    of the particle number spectra   at earlier (upper panels) and later (lower panels) moments of time.  Left panels: $g=1$, Middle panels: $g=2$, Right panels: $g=4$}
\end{figure*}

\begin{figure*}
\includegraphics[width= 0.99 \textwidth]{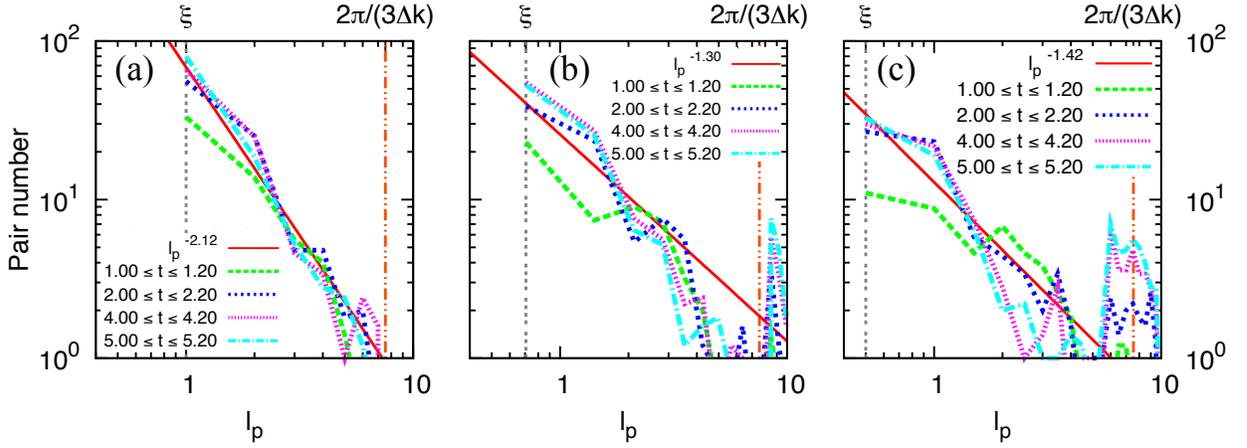}
\caption{\label{f:7} (Color online) Averaged vortex pair number as the function of intervortex length $l_p$. Left panels: $g=1$, Middle panels: $g=2$, Right panels: $g=4$}
\end{figure*}

\begin{figure*}
\includegraphics[width= 0.99 \textwidth]{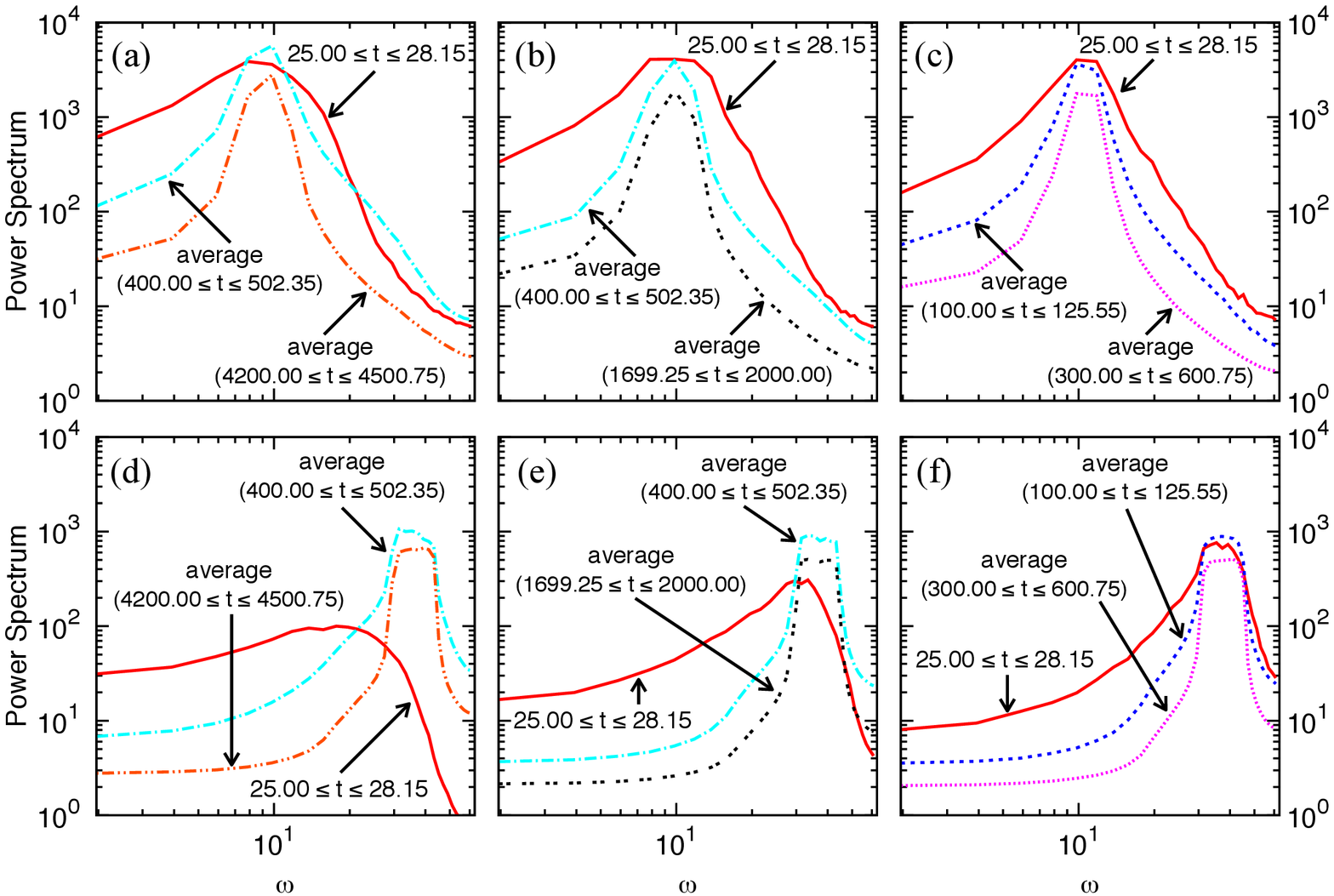}

\caption{\label{f:8} (Color online) Log--log plots
    of the  power spectra for the compressible velocity component with $k=3 \sqrt 2$ (upper panels) and $k=6 \sqrt 2$ (lower panels). Left panels: $g=1$, Middle panels: $g=2$, Right panels: $g=4$}
\end{figure*}

\begin{figure*}\hskip -0.9cm 
\includegraphics[width= 1.04 \textwidth]{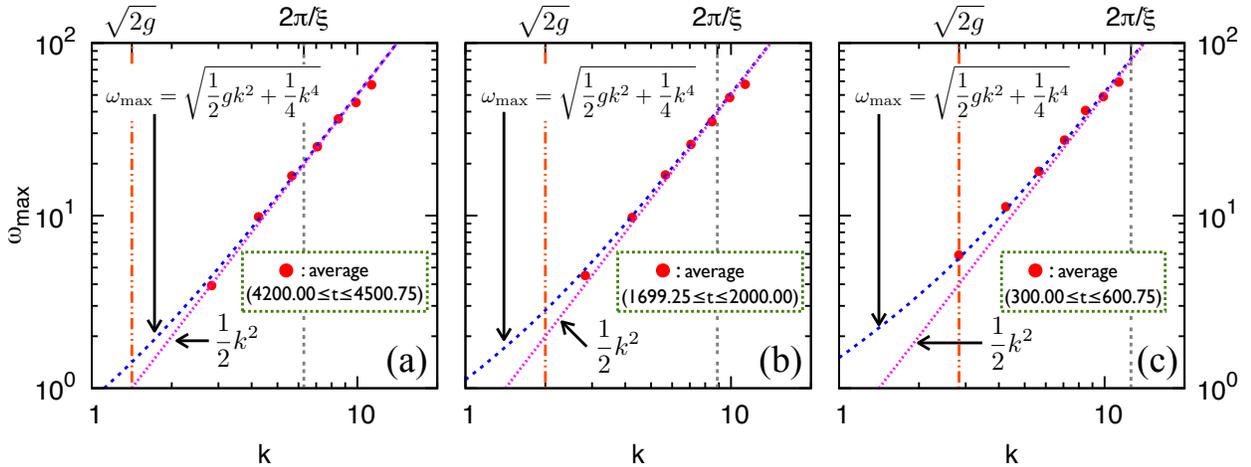}

\caption{\label{f:9} (Color online)  Position of the maximum frequency--power spectra of the compressible velocity component for different wave-vectors, averaged over a long time in the state of full thermodynamic equilibrium (full dots). Bogoliubov's frequency spectrum $\omega\sb{max}(k)$, \Eq{Bog} and its large $k$ asymptotic $\omega\sb{max}(k)\propto k^2$ are also shown. Left panel: $g=1$, Middle panel: $g=2$, Right panel: $g=4$.}
\end{figure*}

\begin{figure*}
\includegraphics[width= 0.99 \textwidth]{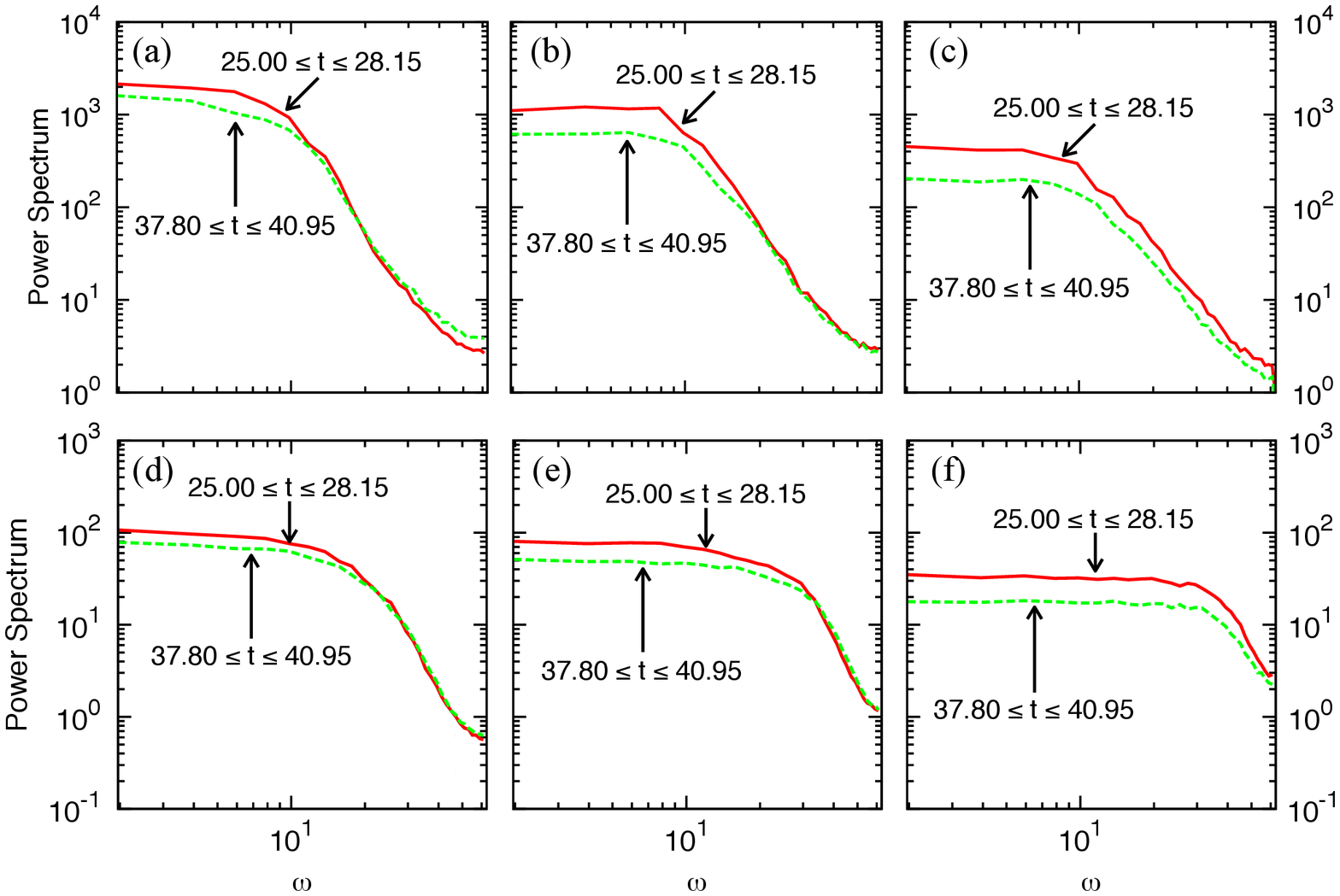}
\caption{\label{f:10} (Color online) Log--log plots
    of the  power spectra for the incompressible velocity component with $k=3 \sqrt 2$ (upper panels) and $k=6 \sqrt 2$ (lower panels). Left panels: $g=1$, Middle panels: $g=2$, Right panels: $g=4$}
\end{figure*}

To investigate the character of the 2D QT we evolve the GPE~\eq{GPE} as described above and compute the following:
\begin{enumerate}
\item Time evolution of all four total energy components~\eq{en} per total particle number $N$ and total vortex number $\C N_{\rm qv}$ for different values of the coupling constant $g=1\,, \ 2\,, \ 4$, shown in \Fig{f:1}.

\item Time evolution of the dimensionless ``level of nonlinearity", defined as $E\sb{int}/(E\sb{kin}+ E\sb{qnt})$, shown in \Fig{f:2}.

\item  Compressible  and incompressible   kinetic energy spectra, $E\sb{kin}\sp c(k)$ and    $E\sb{kin}\sp i(k)$, shown for $g=1\,, \ 2\,, \ 4$ in \Figs{f:3}, \ref{f:4}, and  defined below:
\BSE{spec}\begin{eqnarray}\label{specA}
E\sb{kin}\sp c(t) &=& \frac 1{2N} \int |\tilde{\bm w}\sp c(\bm k)|^2\, d  \bm k\equiv \int \ E\sb{kin}\sp c(k)\, dk\,,~~~~~\\ \label{specB}
E\sb{kin}\sp i(t) &=& \frac 1{2N} \int |\tilde{\bm w}\sp i(\bm k)|^2\, d \bm k \equiv \int \ E\sb{kin}\sp i(k)\, dk\,,\\ \label{specC}
1 &=& \frac 1N\int |\tilde{\Psi}(\bm k)|^2\, d \bm k \equiv \int \ N (k)\, dk\,,
\end {eqnarray}
where
\begin{eqnarray}
\tilde{w}\sp c_{\alpha}(\bm k) &=& \sum_{\beta = 1,2}^{} \frac{k_{\alpha} k_{\beta}}{k^2} \tilde{w}_{\beta}(\bm k)\, ,\\
\tilde{w}\sp i_{\alpha}(\bm k) &=& \sum_{\beta = 1,2}^{} \biggl(\delta_{\alpha \beta}-\frac{k_{\alpha} k_{\beta}}{k^2}\biggr) \tilde{w}_{\beta}(\bm k)\ .
\end {eqnarray}\ese

\item The spectra $E\sb{kin}\sp c(k)$,     $E\sb{kin}\sp i(k)$  and $N(k)$  for different moments of time and different values of $g$, 
shown in \Figs{f:3}, \ref{f:4},  and \ref{f:5} respectively.

\item Mean incompressible kinetic energy flux in $k$-space, $\varepsilon\sp i (k)$ shown in \Fig{f:6} ($g=1\,, \ 2\,, \ 4$).

\item  Distribution of vortex-pair  separations, shown for $g=1\,, \ 2\,, \ 4$ in \Fig{f:7}.

\item Frequency spectra of compressible and incompressible velocity components for different $k$ and different $g$ at later time moments, shown in \Figs{f:8} and \ref{f:10} respectively. Positions of the maxima of the compressible frequency spectra $\omega\sb{max}(k)$ are shown in \Fig{f:9} for different $g$ and different moments of time.

\end{enumerate}

\subsection{\label{ss:flux}Incompressible kinetic energy flux}
One of our aims is to estimate from numerics the value and direction of the energy flux over a variety of scales in the incompressible and compressible subsystems, $\varepsilon\sp i(k,t)$ and $\varepsilon  \sp c(k,t)$.  To do this we use the observation (see below \Fig{f:1}) that during most of the evolution (say for $t>2.5$) the total kinetic energy is practically conserved. We can then neglect the energy exchange between these subsystems. In contrast, there is a permanent energy flux from the incompressible
to the compressible subsystem and this exchange has to be accounted for. Therefore the  (approximate)
balance equations  for the energy densities $E\sb{kin}\sp c(k,t)$ and $E\sb{kin}\sp i(k,t)$, introduced by \Eqs{spec}, can be written as follows:
\BSE{BUD}
\BEA{BUDi}
\frac{\p E\sb{kin}\sp c(k,t)}{\p t}+ \frac{\p \varepsilon\sp c(k,t)}{\p  k} + F\sp{ci}(k,t)&=&0\,, \\ \label{BUDc}
\frac{\p E\sb{kin}\sp i(k,t)}{\p t}+ \frac{\p \varepsilon\sp i(k,t)}{\p  k} - F\sp{ci}(k,t)&=&0\ .
\eea\ese
Here $ F\sp{ci}(k,t)$ describes the energy exchange between subsystems.  Integrating these equations we conclude:
\BSE{BUD}
\BEA{BUDi}
 \varepsilon\sp c(k,t) &= & - \int \limits _{k\sb {min}} ^k \Big [ \frac{\p E\sb{kin}\sp c(k',t)}{\p t} + F\sp{ci}(k',t) \Big ] dk '\,, \\ \label{BUDc}
 \varepsilon\sp i(k,t) & = & - \int \limits _{k\sb {min}}^k \Big [ \frac{\p E\sb{kin}\sp i(k',t)}{\p t} - F\sp{ci}(k',t) \Big ] dk '\ . \eea\ese
A zero-order approximation is applied in the analysis of \Eqs{BUD}  to eliminate $F\sp{ci}(k,t)$ in these equations.  This approximation can be improved as follows. We know that energy flux terms do not contribute to the total balance:
 \BSE{BUD1}
\BEA{BUD1i}
- \int \limits _{k\sb {min}} ^{k\sb {max}}\Big [ \frac{\p E\sb{kin}\sp c(k',t)}{\p t} + F\sp{ci}(k',t) \Big ] dk ' &= & 0\,, \\ \label{BUD1c}
- \int \limits _{k\sb {min}} ^{k\sb {max}}\Big [ \frac{\p E\sb{kin}\sp i(k',t)}{\p t} - F\sp{ci}(k',t) \Big ] dk ' &= & 0\,,  \eea\ese
 which allows us to find from numerics $\int  _{k\sb {min}} ^{k\sb {max}} F\sp{ci}(k',t)  dk '$ at any instant of time, for example, in the  following way:
\BSE{try}
\BE{tryA}\int\limits   _{k\sb {min}} ^{k\sb {max}} F\sp{ci}(k',t)  dk' =   \int \limits _{k\sb {min}} ^{k\sb {max}} \frac{\p \big[ E\sb{kin}\sp i(k',t)- E\sb{kin}\sp c(k',t)\big] }{2\,  \p t}d k'\ .
\ee

\begin{figure*}
\includegraphics[width= 0.99 \textwidth]{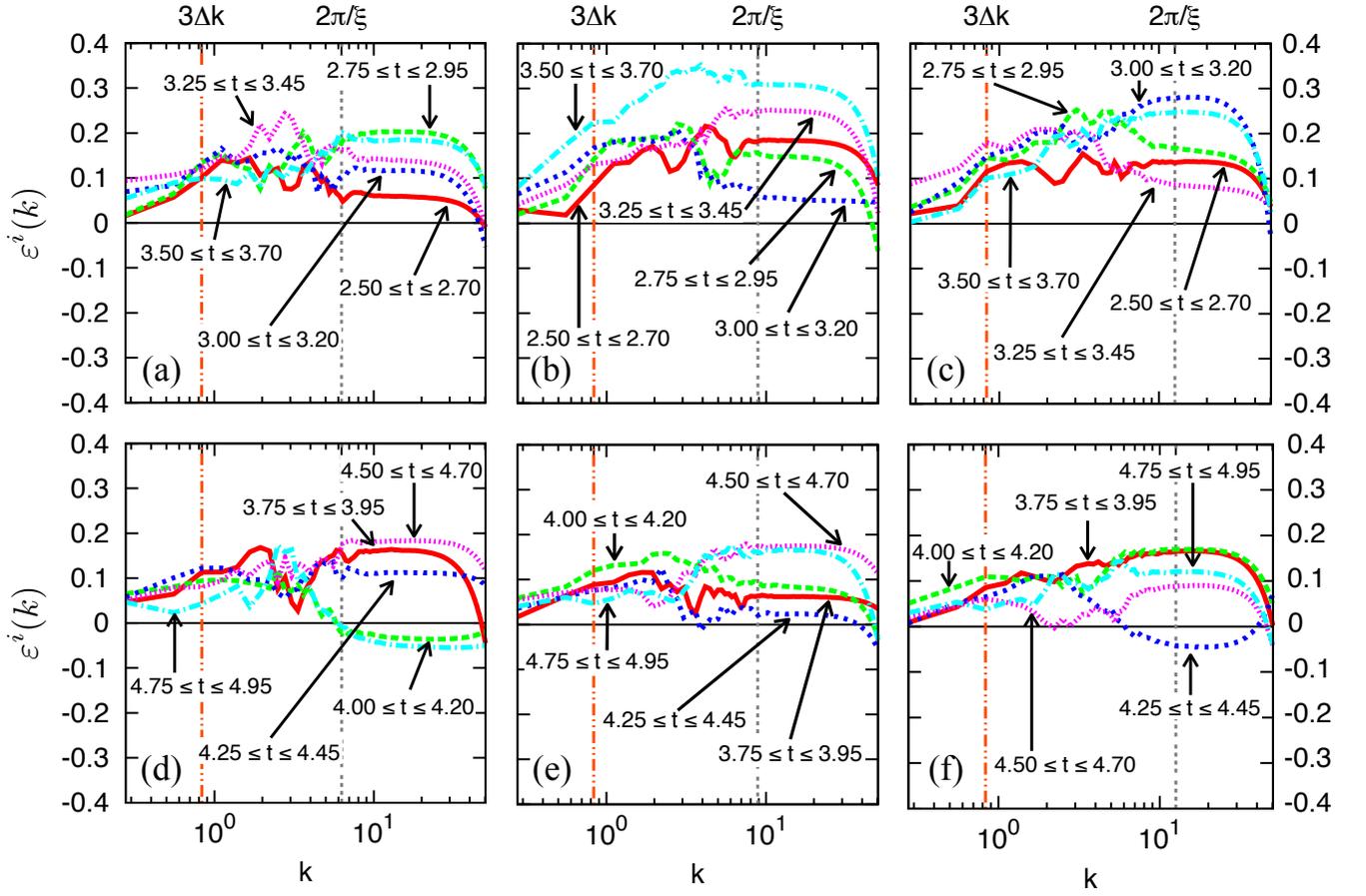}

\caption{\label{f:6} (Color online)  Averaged incompressible kinetic energy flux in a short time interval $\tau=0.20$ in the time range $2.50 \le t \le 4.95$. Left panels: $g=1$, Middle panels: $g=2$, Right panels: $g=4$}
\end{figure*}

Next we can approximate $ F\sp{ci}(k',t)$ in the factorized form
\BE{tryB} F\sp{ci}(k', t) \Rightarrow
f(t)\, \varphi  (k') \,,
 \ee %%
 with a prescribed $k'$-dependent function $ \varphi  (k')$ that can be determined by reasonable physical arguments. For example, assuming  that $ F\sp{ci}(k',t) $ mostly originates from the collapse of vortex pairs, noting that the velocity around  each vortex decays like $1/r'$, giving in the $k'$-representation proportional to $k'$
and that $E_{\rm kin}^{\rm i}(k')\propto k'\, |\tilde{\B w}(k')|^2$. In our calculations we took  $\varphi  (k')\propto k'^3$, or in the normalized form
\BE{tryC}
\varphi  (k')=4 k'^3/k\sb{max}^4\ .
\ee
Then from \Eq{tryA} we find
\BE{tryD}f(t) \approx    \int \limits _{k\sb {min}} ^{k\sb {max}} \frac{\p \big[ E\sb{kin}\sp i(k',t)- E\sb{kin}\sp c(k',t)\big] }{2\,  \p t}d k'\ .
\ee
\ese
Equations~\eq{tryB}, \eq{tryC}, and \eq{tryD} can be substituted back into \Eqs{BUD} to get
an improved  approximation for the flux.
\label{Ryu} Finally, we obtain the incompressible kinetic energy flux in a form with (\ref{tryD}):
\BE{i:flux}
\varepsilon\sp{\rm i} (k, t) \approx -\int \limits _{k\sb {min}} ^k \frac{\p E_{\rm kin}^{\rm i}(k^\prime, t)}{\p t} d k^\prime + f(t) \Bigg( \frac k{k\sb{\rm max}}\Bigg)^4.
\ee
Numerical results for the energy flux, obtained in this way, are shown in \Fig{f:6} and will be discussed later.

%%%%%%%%%%%%%%%%%%%%%%%%%%%%%%%%%%%%%%%%%%%%%%%%%%%%%%

%%%%%%%%%%%
%%%%%%%%%%%%

\section{\label{s:GPE}Decay of 2D compressible turbulence}

\subsection{\label{ss:free}Free evolution of system energies and vortex number}
Energy evolutions of the GP system, starting from the same initial condition $\Psi(\B r, 0)$, but with different values of the coupling constant, $g=1$, $g=2$, and $g=4$, are shown in \Fig{f:1}. Visual analysis of the figure leads to several qualitative statements concerning the system behavior:

\textbullet~  As expected the total energy, Hamiltonian is time independent and has larger values for larger $g$ (due to the $g |\Psi|^4$ contribution).

\textbullet~  At the initial stages of the time evolution (for $t \lesssim 5$, upper panels) an intensive process of vortex creation leads to fast transformation of the compressible  kinetic energy into incompressible kinetic energy and quantum energy (not shown). 
The incompressible kinetic energy is larger than the compressible kinetic energy for the time interval $2 \lesssim t \lesssim 4$. 

\textbullet~   At later stages ($t \gtrsim 5$) the total kinetic energy  is practically independent of time. Moreover its values  are more or less the same for all $g$. This can be easily understood: the kinetic energy is quadratic in $\Psi$ and  does not contain the nonlinear term proportional to  $g$.

\textbullet~  The largest maximal value of the vortex number $\C N_{\rm qv}(t)$ is achieved at a smaller value $g=1$ (Left panel) at the crossover time $t\approx 5$.  This is because for $t \gtrsim 5$ the probability of the dominant \emph{nonlinear process} of vortex-pair annihilation is larger for larger $g$.  We see that the fastest decay of $\C N_{\rm qv}(t)$ corresponds to $g=4$, Right panel.

\textbullet~  A faster decay of the vortex number for larger $g$ leads to a larger increase in the flow compressibility with an increase in $g$. Correspondingly, the most compressible flow (largest value of the ratio $E\sb {kin}\sp c/ E\sb {kin}\sp i$)
is reached (in the presented data) for $g=4$, and the least for $g=1$.

\label{Ryu}\textbullet~  The system finally reaches its equilibrium state without quantized vortices for all $g$. It takes longer to reach thermodynamic equilibrium state for smaller $g$.

\textbullet~  To clarify in more detail why a large value of the coupling $g$ corresponds to a faster time evolution, we compute and plot in \Fig{f:2} the time evolution of the interaction energy $E\sb{int}$, defined by \Eq{E-int} (and proportional to $g$). This energy has to be compared with the sum of the kinetic and quantum energies, $E\sb{kin}+E\sb{qnt}$,  which originates from the quadratic term in the Hamiltonian~(\ref{H}) or the linear term in the GPE~\eq{GPE}. This sum is shown in \Fig{f:2}. We see that typically $E\sb{int}$ is essentially smaller than $E\sb{kin}+E\sb{qnt}$ and decays in time. Therefore the ratio
\be\label{nl}
R\= E\sb{int}\big / (E\sb{kin}+E\sb{qnt}),
\ee
shown in \Fig{f:2}, characterizes how far our system deviates from its linear  approximation. Accordingly, the ratio $R$ can be called the \emph{nonlinearity  level}. 
The larger $R$, the more quickly the system should  evolve to its equilibrium state. 
Indeed, we see that the maximal value of $R\simeq 0.8$ is reached (at $t\approx 0.8$) for $g=4$, while for $g=1$ the maximal is $R\approx 0.2$, as expected. An important point is that $R$ is essentially less than unity (e.g. $R\sim 0.2$ for $g=4$ is the highest nonlinearity level at times $1.5 <t < 5$) and decreases with time. This means that in an analytical approach  to  the problem, the interaction can be treated through a perturbation approach with respect of $R$.

\subsection{\label{ss:power}Evolution of the compressible and incompressible kinetic energy  spectra}
The energy distribution between scales (energy spectra) for   $g=1,\,   2,\,  4$ and at different moments of time during free decay are shown in \Fig{f:3} for the compressible subsystem and in \Fig{f:4} for the incompressible subsystem. We conclude that:

\textbullet~
At earlier stages (upper panels in both \Figs{f:3} and \ref{f:4}) the compressible and incompressible kinetic energies, being  initially located at small $k \le 3\Delta k$, propagate toward larger $k$, still (almost) monotonically decaying toward large $k$.

\textbullet~
The intermediate moments of time (around $t\sim 3$) will be discussed in detail in the next \Sec{ss:2D-direct}. Here we only note that the incompressible spectra  are  close to the Kolmogorov--Obukhov law $E_{\rm kin}^{\rm i}(k)\propto k^{-5/3}$ shown in the upper panels of \Fig{f:4}. This behavior is related to the energy cascade.

\textbullet~
At later times the energy begins to accumulate at large $k$ and the energy spectra asymptotically approach the quasi-stationary state (practically coinciding e.g. for $t=15$ and $t=40$ within the error bars, related to the lack of statistical averaging).

\textbullet~
The compressible energy spectra $E\sb {kin} \sp c (k)$ at this stage are close to the thermodynamic equilibrium with the energy equipartition between degrees of freedom. In 2D systems this gives (for one-dimensional spectra)  $E\sb {kin} \sp c (k)\propto k$, as shown in the lower panels in \Fig{f:3}. This state will be discussed in more detail below in \Sec{ss:frec}. 

\textbullet~
The achieved (quasi-stationary) distribution of the incompressible kinetic energy $E\sb {kin} \sp i$ (lower panels in \Fig{f:4}) \emph{does not} correspond to the thermodynamic equilibrium, as we can na\"{\i}vely expect. This is related to the fact that at these times the system \emph{does not achieve full equilibrium}.
Indeed, the lower panels in \Fig{f:1}  show that the incompressible kinetic energy continues  to converge to the compressible energy. We interpret this stage as  a kind of flux (not thermodynamic) equilibrium, when the spectrum $E\sb {kin} \sp c (k)$ is determined by the energy flux from the incompressible to the compressible subsystem. We think that at these times the energy exchange within these subsystems (that should lead to thermodynamic equilibrium) plays a subdominant role.

\label{Ryu}\textbullet~
After a long time evolution, all the kinetic energy finally comes to comprise the compressible component. This is the full thermodynamic state. We show the long-time averaged compressible kinetic energy spectra $E_{\rm kin}^{\rm c}(k)$. As expected, this is proportional to $k$.

\subsection{\label{ss:part}Evolution of the particle number spectra}
As we can see from \Eq{N}, the total particle number  is quadratic in $\Psi(\B r)$ and therefore can be presented in the $\B k$-representation by Eq.~(\ref{specC}), shown in \Fig{f:5} for different time intervals and all values of $g=1\,, \ 2$ and 4. As we see,  initially, $N(k)$ is  localized in small-$k$.  Through the turbulent state, the spectrum asymptotically approaches the form $N(k) \propto k^{-1}$ in the wavenumber range $k \lesssim 2\pi/\xi$.  To rationalize this behavior, we note that in the full thermodynamic equilibrium   this dependence should be related to $E\sb{\rm kin}\sp{\rm c}(k) \propto k$. Indeed, in this state, the nonlinearity level Eq.~\eq{nl} is small, meaning that  the interaction energy can be neglected in comparison with  the kinetic and quantum energy. In addition, the quantum energy which cannot be considered negligible corresponds to zero-point motion and this term does not give rise to particle currents\cite{Peth-02}. Thus, only the kinetic energy spectrum is related to the particle spectrum. Moreover,  almost all  the kinetic energy becomes the compressible component and there are no vortices, and as a result $\theta$ has no singularities and becomes of order unity.
Thus the  kinetic energy density can be estimated using a wavenumber $k$ as follows:
\BE{estim1}
\C E \sb{kin} \simeq \C E \sb{kin}^{\rm c} \simeq \frac 12 \rho |\B \nabla \theta|^2 \sim \frac 12 \rho k^2 |\theta|^2 \sim \frac 12 \rho k^2,
\ee
which becomes in $\B k$-space using (\ref{specA}) and (\ref{specC}), $\frac 12 |\tilde{\B w} (\B k)|^2 \simeq \frac 12 k^2|\tilde \Psi(\B k)|^2$, as a result,
\BEA{estim3}
E\sb{\rm kin}\sp{\rm c}(k) &\simeq& \frac 12 k^2N(k).
\eea
In this way, the two relations $E\sb{\rm kin}\sp{\rm c}(k) \propto k$ and $N(k) \propto k^{-1}$ hold simultaneously. This relation is similar to the relation between enstrophy $\Omega$ and kinetic energy  $E$ in 2D classical fluids.

\subsection{\label{ss:2D-direct}Observation of 2D-direct energy cascade}

Returning to the   discussion of  \Figs{f:1} and \ref{f:3}, we note the following:

%\begin{description}

\textbullet~
The incompressible energy spectra at these times can be interpreted as closed to the classical Kolmogorov--Obukhov distribution $E\sb {kin} \sp i(k)\propto k^{-5/3}$, shown in the upper panels of \Fig{f:4}.
%This correspondence is clearer for $g=2$. For smaller $g$ the compressibility is small and essentially destroys the enstrophy invariant that forbids     the direct energy cascade. For larger $g$ the incompressible part of the energy is small enough to be considered independent of the motion in the compressible part.

\textbullet~
Bearing in mind that during the system evolution the kinetic energy clearly propagates from small $k$ (where it was initially located in $E_{\rm kin}^{\rm i}(k)$) towards large $k$ we conclude that at some region of the  system parameters we can observe a \emph{2D-direct energy cascade}, which has previously been observed only in 3D hydrodynamic systems.

\textbullet~
Numerical analysis of the incompressible kinetic energy flux supports the above explanation. Using \Eq{i:flux}, we calculate the incompressible kinetic energy flux. Averaged data for a short time interval $\tau = 0.20$ with five sets of data are shown in Fig.\ref{f:6}. The flux $\varepsilon^{\rm i}(k)$ takes positive values for $3\Delta k \lesssim k \lesssim 2\pi/\xi$ at least for $2.50 \le t \le 4.95$. This strongly supports the occurrence of a 2D direct energy cascade. Moreover, owing to the second term in Eq.~(\ref{i:flux}), especially effective for high wavenumbers, $\varepsilon^{\rm i}(k)$ tends to vanish at $k_{\rm max}$.

\textbullet~
The Kolmogorov energy spectrum $E_{\rm kin} ^{\rm i}(k)\propto k^{-5/3}$ was previously observed in decaying 2D turbulence in the GPE~\cite{Par-05}. However, this observation was interpreted in terms of the known 2D turbulence inverse energy cascade, predicted by Kraichnan~\cite{Kra-67}.  We think that this interpretation is mistaken. In particular, if this interpretation is true, this spectrum has to be followed by a $k^{-3}$ spectrum for a direct enstrophy cascade according to Kraichnan's scenario. However, the authors of Ref.~\cite{Par-05} ``observe a $k^{-6}$ dependence in this range", instead of $k^{-3}$ dependence.

\textbullet~  When the Kolmogorov--Obukhov spectrum is formed, a ``two-dimensional'' Richardson cascade can be seen. We can identify the cores of the quantized vortices by finding the phase defects of the wave function. Then, we choose the shortest intervortex pair length $l_p$ for all vortices. In \Fig{f:7}, the averaged vortex pair number $\C N_{\rm pair} (l_p)$ is shown and is proportional to $l_p ^{-n}$, where $n$ depends on $g$ and $1.30 \le n \le 2.12$. This power law suggests a self-similar spatial structure. For two-dimensional vortices, one of the most effective lengths, corresponding to the length of three-dimensional vortex ring, is the intervortex length. We conclude that \Fig{f:7} implies the following: First, vortex pairs with distances comparable to $2\pi/(3\Delta k)$ are created. Second, the vortex pair distance progressively decreases. Finally, when the vortex pair distance becomes comparable to $\xi$, the pair is annihilated.

\subsection{\label{ss:frec}Frequency power spectra and types of motions}
\subsubsection{Compressible type of motion}
Important information about types of motion can be extracted from the frequency power spectrum, which is the Fourier transform of the different-time pair correlation function of the motion amplitude. For example, if an amplitude $A(t)$ oscillates with a particular frequency $\omega_0$, i.e.  $A(t)=A_0 \exp (i\omega_0 t)$, then 
$$J(\tau)\=\< A(t+\tau)A^*(t)\>=A_0^2 \exp (i \omega_0\tau)\,, $$ giving 
$$\~ J(\omega)= \int J(\tau)\exp (- i\omega \tau)d \tau\propto \delta (\omega-\omega_0)\ .$$
In other words, pure periodic motion with frequency $\omega_0$ has a frequency power spectrum with a very intensive peak of zero width at $\omega=\omega_0$. It is easy to check that decaying oscillations  $A(t)=A_0 \exp (-\gamma t + i\omega_0 t)$ have the power spectrum
\be\label{ps}
\~ J(\omega)\propto \frac {\gamma}{(\omega-\omega_0)^2+\gamma^2}\,,
\ee
i.e. a peak at frequency $\omega_0$ and width $\gamma$, inversely proportional to the life time of the motion. 

We observed exactly this kind of frequency power spectra for the compressible velocity component at  later times, as shown in Fig.\ref{f:8} for $k=3\sqrt 2$ and $k=6\sqrt 2$.  As the system evolves to the thermodynamic equilibrium state, the power spectrum forms sharp peaks. This tendency is also seen in the frequency power spectrum of the wave function $\Psi$ (not shown). We repeated this analysis for different $k$ and in \Fig{f:9} we plot the position of the maxima of the frequency power spectra of compressible motion, averaged over a long time interval in the thermodynamic equilibrium state. Bearing in mind that for these periods the nonlinearity level $R$ is very small (about 0.04 for $g=4$ and even smaller for $g=2$ and 1), so one can consider these fluctuations as almost linear perturbations on the background of the rest state. The Eigenfrequencies of these oscillations have been found by   Bogoliubov~\cite{Peth-02} with the result:
\BE{Bog}
\omega\sb{\rm max} = \sqrt{ \frac 12 g k^2 + \frac 14 k^4 }\,,
\ee
plotted in \Fig{f:9}. The excellent agreement between the theoretical and numerical results indicates that the observed thermodynamical fluctuation of the compressible velocity component is indeed  Bogoliubov's elementary excitations. The relatively small, but finite width of the observed peaks characterizes the finiteness of the life time of these fluctuations, caused by interaction of the fluctuations with different $k$-vectors.

\subsubsection{Incompressible type of motion}
What kind of frequency power spectrum can we expect for  the incompressible type of motion? To answer this we note that an incompressible fluid exhibits vortex motions in which the life time and turnover time are of the same order. This may mean that the position of the peak maximum $\omega_0$ and its width $\gamma$ must be of the same order of magnitude. Moreover, more detailed analysis shows that due to symmetry, $\omega_0\=0$. In fully developed turbulence, the Kolmogorov-41 dimensional estimates gives $\tau^{-1}(k)\simeq \varepsilon ^{1/3} k^{2/3}$ for the inverse life-time of vortices, where $\varepsilon$ is the energy flux.

Under a trivial consideration we can expect $\gamma\simeq \tau^{-1}(k)$. However, this will be true only in the reference system, co-moving with the small vortices. In the laboratory  reference system, where the mean velocity of the fluid is absent, we have to account for the sweeping of small vortices in the velocity field $\B V$ of the largest vortices. This gives a Doppler shift of the peak position equal to $\B k \cdot \B V$. Noting that the turbulent velocity field is random, we average the shifted frequency peaks over the statistics of large-scale motions, in which positive and negative velocities have the same  probability.   As a result we do not shift the peak, but instead widen it to $\gamma(k)\simeq \sqrt {\< V^2 \>}\, k$. 

Qualitatively, this is what is seen in  \Fig{f:10}, where  the frequency power spectrum of the incompressible effective velocity component with $k=3\sqrt 2$ and $k=6\sqrt 2$ are shown. As expected, they do not form sharp peaks. Also the peak width for $k=3\sqrt {2}$ (upper panels)  are smaller than for  $k=6\sqrt {2}$, (lower panels).  We consider these facts as additional support for the fact that incompressible motion demonstrates turbulent vortex behavior. 

As we already demonstrated,  $E\sb{\rm kin}\sp{\rm i}(t)$ approaches $0$ after $t \lesssim 5$, taking a long time, and all the vortices finally entirely disappear. As a result, the compressible effective velocity component dominates in the thermodynamic equilibrium state.

\section{\label{s:concl}Conclusion}
%\Fbox{Ryu, pls eliminate all abbreviations below for a general reader who wants to read only Concussion and is not familiar with them.}

%\Fbox{Our  results show how clever we are. The rest will  be written}
In this paper we predict and numerically demonstrate a two-dimensional direct energy cascade with the Kolmogorov--Obukhov's $-5/3$ law in two-dimensional quantum turbulence with the two-dimensional Gross--Pitaevskii model. The cascade is mainly caused by the \emph{compressibility} of Bose--Einstein condensates which forbids enstrophy conservation. A two-dimensional direct energy cascade has not previously been observed in typical two-dimensional classical turbulence in the decaying case, which features direct enstrophy cascade and self-organization of the system, because of the conservation laws of both energy and enstrophy.

In addition, we demonstrate that the turbulent state shifts to the thermodynamic equilibrium state full of elementary excitations without quantized vortices. The dispersion relation calculated from the frequency--power spectrum of the compressible velocity components shows that the equilibrium state is filled with Bogoliubov excitations.

It is important to note that two-dimensional quantum turbulence is not restricted to theory. It is possible to design experiments of two-dimensional quantum turbulence in pseudo two-dimensional atomic Bose--Einstein condensates, superfluid $^4$He, and superfluid $^3$He. So far, almost all studies of quantum turbulence have concentrated on the three-dimensional case, except for a few works~\cite{Par-05,Gou-09}. However, at least in atomic Bose--Einstein condensates, the compressibility causes a significant change from two-dimensional classical turbulence, and in a certain parameter range two-dimensional quantum turbulence differs from two-dimensional classical turbulence. Thus, we are certain that the two-dimensional direct energy cascade in two-dimensional quantum turbulence provides a new phase of study of quantum turbulence.

%\bigskip

\acknowledgements
MT acknowledges the support of a Grant-in Aid for Scientific Research from JSPS (Grant No. 21340104) and a Grant-in Aid for Scientific Research on Priority Areas from MEXT (Grant No. 17071008).
VL acknowledges the kind hospitality of Osaka City University and the support of the  \emph{Japan Society for the Promotion of Science} (grant \# S-09147)  where this project was started
and the partial support of  the \emph{European Community -- Research Infrastructures under the FP7 Capacities Specific Programme, MICROKELVIN} (project \# 228464) and the \emph{U.S. - Israel Binational Science Foundation} (grant \# 2008110).

\end{document}